\documentclass[10pt,twocolumn,pra,showpacs,floatfix]{revtex4}
\usepackage{dcolumn}                 
\newcolumntype{w}[1]{D{.}{.}{#1}}

\newcommand*{\cent}[1]{\multicolumn{1}{c}{$#1$}}
\usepackage{times}
\usepackage{nicefrac}
\usepackage{amsmath}
\usepackage{amsfonts}
\usepackage{amssymb}
\usepackage{amsthm}

\usepackage{graphicx}

\begin{document}
\preprint{Version 1.0}
\title{Applications of four-body exponentially correlated functions}

\author{Mariusz Puchalski}
\email[]{mpuchals@fuw.edu.pl}

\author{Krzysztof Pachucki}
\email[]{krp@fuw.edu.pl}

\affiliation{Institute of Theoretical Physics, University of Warsaw,
             Ho\.{z}a 69, 00-681 Warsaw, Poland}

\begin{abstract}
We demonstrate the applicability of four-body exponentially correlated
functions for the accurate calculations of relativistic effects
in lithium-like atoms and present results for matrix elements
of various operators which involve negative powers of interparticle distances.
\end{abstract}

\maketitle

\section{Introduction}
In the accurate evaluation of atomic energy levels not only the nonrelativistic 
energy, but also relativistic and QED effects have to be calculated with the
high precision. The only approach which consistently accounts for all 
corrections in small atomic systems
is the one based on the expansion of the energy in the fine
structure constant $\alpha$
\begin{equation}
E = m\,\bigl[\alpha^2\,{\cal E}^{(2)} + \alpha^4\,{\cal E}^{(4)} + 
    \alpha^5\,{\cal E}^{(5)} + \alpha^6\,{\cal E}^{(6)} + 
    \alpha^7\,{\cal E}^{(7)} + \ldots\bigr].
\end{equation}
Each term in this expansion can be expressed
as the expectation value of some effective Hamiltonian
with the nonrelativistic wave function. Namely
${\cal E}^{(2)}$ is the nonrelativistic energy (in atomic units),
${\cal E}^{(4)}$ is the relativistic correction, which for states with 
the vanishing angular momentum is the expectation value of $H^{(4)}$ in 
Eq. (\ref{df_H4}). ${\cal E}^{(5)}$ and higher order corrections 
are expressed in terms of matrix elements of some
more complicated operators. They have been calculated for low lying states of
the helium atom and helium-like ions up to the order $m\,\alpha^6$ \cite{pach1},
and for the particularly important case of $2^3P_J$ splitting
up to the order $m\,\alpha^7$ \cite{pach2}. One of the sources of this achievement
was the flexibility of the explicitly correlated exponential basis set,
which due to its correct analytic properties, makes possible accurate evaluation
of matrix elements with complicated and singular operators.

In the case of the lithium atom and light lithium-like ions all corrections up to
${\cal E}^{(5)}$ have been accurately calculated \cite{king_rel, drake,
lit_rel}, but not that of higher orders. 
The principal reason for the much slower progress for three-electron
atoms is the difficulty in handling integrals with explicitly correlated functions.
The commonly used explicitly correlated Gaussian functions do not
have right analytic properties, for example they do not satisfy the cusp condition,
and therefore cannot be used for the calculation of higher order
relativistic corrections, like ${\cal E}^{(6)}$. Hylleraas basis functions
have the right analytic behavior: the accuracy in solving the Schr\"odinger 
equation is the highest among all other basis functions,
but it is difficult to handle Hylleraas integrals involving quadratic
negative powers of two different interparticle distances. 
Such integrals appear in the evaluation of  ${\cal E}^{(6)}$ and     
for this reason other basis functions have been investigated
in the literature.

Zotev and Rebane in \cite{rebane_ham} were the first to apply exponentially correlated functions
\begin{equation}
\phi(\vec r_1, \vec r_2, \vec r_3) =
e^{-\alpha_1\,r_1-\alpha_2\,r_2-\alpha_3\,r_3-\beta_1\,r_{23}-\beta_2\,r_{13}-\beta_3\,r_{12}}\,,
\label{df_phi}
\end{equation}
in variational calculations for Ps$_2$ and the other exotic molecules.
They have found a simplified formula for matrix elements
of the nonrelativistic Hamiltonian and presented numerical results
of variational calculations with a few basis functions.
In our recent paper \cite{slater1} we presented an efficient algorithm 
for the evaluation of integrals involving powers of $r_{i}$ and $r_{ij}$
\begin{eqnarray}
g(n_1,n_2,n_3,n_4,n_5,n_6) &=& \int \frac{d^3 r_1}{4\,\pi}\,
                               \int \frac{d^3 r_2}{4\,\pi}\,
                               \int \frac{d^3 r_3}{4\,\pi}\, \nonumber \\
&& \hspace{-1.5cm} r_{1}^{n_1-1}\,r_{2}^{n_2-1}\,r_{3}^{n_3-1}\,
r_{23}^{n_4-1}\,r_{31}^{n_5-1}\,r_{12}^{n_6-1}\nonumber \\ 
&& \hspace{-1.5cm} e^{-w_1\,r_1-w_2\,r_2-w_3\,r_3-u_1\,r_{23}-u_2\,r_{13}-u_3\,r_{12}}  \nonumber \\
\label{df_g}
\end{eqnarray}
with $n_a$ being nonnegative integers. It is based on recursion
relations which start from the master Fromm-Hill integral
\cite{fromm,harris_slater}, where all $n_a=0$. 
We have applied this algorithm to the variational calculations of 
the ground state of Li and Be$^+$ with up to 128 functions. The comparison
of nonrelativistic energies with the ones obtained with much larger number of Hylleraas functions  
indicates that the exponential representation of the three-electron wave function is very efficient.

The class of integrals in Eq. (\ref{df_g}) with nonnegative $n_a$ is sufficient for nonrelativistic 
energies \cite{slater1,rebane_ham}.  However, 
it is not sufficient to calculate the leading relativistic effects described 
by Breit-Pauli Hamiltonian, which for S-states takes the form
\begin{eqnarray}
H^{(4)} &=&
\sum_a \biggl\{-\frac{\vec p^{\,4}_a}{8\,m^3} +
\frac{ \pi\,Z\,\alpha}{2\,m^2}\,\delta^3(r_a)\biggr\}
\nonumber \\
&& \hspace{-1cm} +\sum_{a>b} \biggl\{
\frac{ \pi\,\alpha}{m^2}\, \delta^3(r_{ab})
-\frac{\alpha}{2\,m^2}\, p_a^i\,
\biggl(\frac{\delta^{ij}}{r_{ab}}+\frac{r^i_{ab}\,r^j_{ab}}{r^3_{ab}}
\biggr)\, p_b^j \biggr\}. \nonumber \\
\label{df_H4}
\end{eqnarray} 
Its matrix elements involve an extended class of integrals with exactly one of $n_a$ equal to $-1$,
and all others are nonnegative, while that for leading QED effects involve integrals with $n_a=-2$.
This is well known from calculations with Hylleraas basis functions, where all
$u_a$ in Eq. (\ref{df_g}) are equal to zero. 
Hylleraas extended integrals of that kind have been extensively studied in 
\cite{king_ext1,king_ext2,luchow_ext,porras_ext,yan_ext,king_ext3,feldmann_ext,pelzl_ext} 
using multipole-type of expansions 
and recently by present authors using analytical recurrence approach \cite{rec_ext,lit_wave}. 
Both methods had been successfully applied in high-precision calculations of leading relativistic 
and QED corrections to the energy of lithiumlike systems \cite{king_rel,yan_rel,lit_wave}. 
There are no similar studies for exponentially correlated integrals to the best our knowledge, 
and for the first time we present them in this work.

In the calculation of relativistic and QED effects beyond leading order, 
${\cal E}^{(6)}$ for example, 
another class of integrals appears with two quadratic inverse powers of interparticle distances. 
There are only few studies in the literature for three-electron Hylleraas 
integrals \cite{luchow_ext,porras_ext,pelzl_ext}. The algorithm by King \cite{pelzl_ext} 
seems to be too slow for a large scale computation, where integrals with $\Omega = \sum_a n_a$ 
of order 30 have to be performed. The evaluation of these integrals is quite difficult
with the recursion method and have not yet been worked out so far. 

In the case of exponentially correlated integrals the problem seems to be even more severe, 
since the master integral with $u_a\neq 0$ is much more complicated.  
However, being able to optimize nonlinear parameters of each function independly,
one does not need to use large powers of interparticle distances in the basis set.
For S-states it is sufficient to use functions of the form (\ref{df_phi}). In such a case,
having an analytical and thus accurate method for $g(n_1,n_2,n_3,n_4,n_5,n_6)$ with nonnegative $n_a$,
inverse negative powers of the interparticle distance can be obtained by
the numerical integration with respect to the corresponding parameter $w_a$ or $u_a$.
It however requires a good control of numerical accuracy of the master integral 
and of recursion relations in Eq. (\ref{df_rec}).
The usage the higher precision arithmetic is essential in some critical 
areas of the integration. A key feature of our numerical integration  
strategy is the adapted quadrature, 
which allows one to get the high accuracy with a very small number of points. 

We demonstrate our method on examples with expectation value
of various operators on lithium ground state. Results obtained
for matrix elements involving single $n_a=-1$ are compared to the most accurate 
ones obtained with the Hylleraas basis set. 
Good numerical convergence of results for matrix elements involving two negative powers, 
for example $1/(r_a^2 r_b^2), 1/(r_a^2 r_{ab}^2), 1/(r_{ab}^2 r_{bc}^2)$. 
indicate that this integration approach can be used for the 
calculation of higher order relativistic corrections, for example 
$m\,\alpha^6$ and $m\,\alpha^7$ effects in the hyperfine and fine structure 
of lithium-like systems.

\section{Calculation of integrals}

\subsection{Tetrahedral symmetry}
\begin{figure}[htb]
\includegraphics[height=5cm]{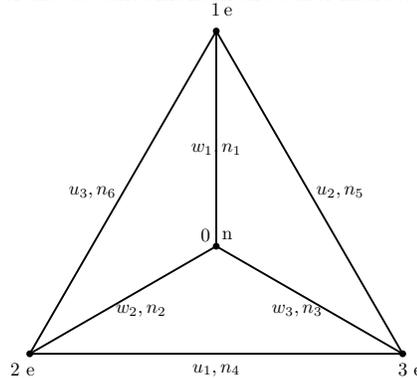}
\caption{Tetrahedron representing the integral from Eq. (\ref{df_phi})}
\label{fig_tetrahedron}
\end{figure}

An important property of integrals defined in Eq.~(\ref{df_g}) 
is the tetrahedral symmetry which is equivalent to
the permutation group $S_4$. We can assign vertices 1,2,3 to 
the electrons and 0 to the nucleus as shown in Fig.~(\ref{fig_tetrahedron}),
and to edges we assign $u_a, w_a$ and $n_a$ parameters of a given integral. 
The symmetry group $S_4$ corresponds to 24 renumbering of vertices 0,1,2, 
and 3, which means also  a relevant change of parameters on the edges.
The generated symmetry relations allow us to reduce the number of 
recurrence formulas for Slater integrals. It is necessary to derive 
only one recurrence scheme, and the formulas for the advancement in
the other indices can then be obtained by application of the $S_4$ symmetry.

\subsection{Basic integrals}

The evaluation method for $g(n_1,n_2,n_3,n_4,n_5,n_6)$ for 
the basis class with all nonnegative $n_a$ was first
presented by Harris in \cite{harris_rec} and later by us in
Ref.~\cite{slater1}. Here we present only a short summary,
which is needed for the evaluation of the extended integrals.
The master integral $g_0\equiv g(0,0,0,0,0,0)$ satisfies 
the following differential equation
\begin{equation}
\sigma\,\frac{\partial g_0}{\partial w_1} +
\frac{1}{2}\,\frac{\partial\sigma}{\partial w_1}\,g_0 + P = 0\,,
\label{diff}
\end{equation}
where the $S_4$ symmetric polynomial $\sigma$ is of the form
\begin{eqnarray}
\sigma &=& u_1^2\,u_2^2\,w_3^2  + u_2^2\,u_3^2\,w_1^2  +
u_1^2\,u_3^2\,w_2^2 + w_1^2\,w_2^2\,w_3^2
\nonumber \\ &&
 + u_1^2\,w_1^2\,(u_1^2+w_1^2-u_2^2-u_3^2-w_2^2-w_3^2)
\nonumber \\ &&
 + u_2^2\,w_2^2\,(u_2^2+w_2^2-u_1^2-u_3^2-w_1^2-w_3^2)
\nonumber \\ &&
 + u_3^2\,w_3^2\,(u_3^2+w_3^2-u_2^2-u_1^2-w_1^2-w_2^2)\,,\label{sigma}
\end{eqnarray}
and function $P$ is given by
\begin{widetext}
\begin{eqnarray}
P &=& -u_1\,w_1\,[(u_1 + w_2)^2 - u_3^2] \,
\Gamma(0,0,-1;u_1+w_2,u_3,u_2+w_1) \nonumber \\ &&
 -u_1\,w_1\,[(u_1 + u_3)^2 - w_2^2]\,\Gamma(0,0,-1;u_1+u_3,w_2,w_1+w_3)
\nonumber \\ &&
 +[u_1^2\,w_1^2 + u_2^2\,w_2^2 - u_3^2\,w_3^2 + w_1\,w_2\,(u_1^2 + u_2^2 - w_3^2)]\,\Gamma(0,0,-1;w_1+w_2,w_3,u_1+u_2)
\nonumber \\ &&
 +[u_1^2\,w_1^2 - u_2^2\,w_2^2 + u_3^2\,w_3^2 +
w_1\,w_3\,(u_1^2 + u_3^2 -
w_2^2)]\,\Gamma(0,0,-1;w_1+w_3,w_2,u_1+u_3)
 \nonumber \\ &&
 -[u_2\,(u_2 + w_1)\,(u_1^2 + u_3^2 - w_2^2) - u_3^2\,(u_1^2 + u_2^2 - w_3^2)]\,\Gamma(0,0,-1;u_2+w_1,u_3,u_1+w_2)
\nonumber \\ &&
 -[u_3\,(u_3 + w_1)\,(u_1^2 + u_2^2 - w_3^2) - u_2^2\,(u_1^2 + u_3^2 - w_2^2)]\,\Gamma(0,0,-1;u_3+w_1,u_2,u_1+w_3)
\nonumber \\ &&
 + w_1\,[w_2\,(u_1^2 - u_2^2 + w_3^2) + w_3\,(u_1^2 + w_2^2 - u_3^2) ]\,\Gamma(0,0,-1;w_2+w_3,w_1,u_2+u_3)
\nonumber \\ &&
 +w_1\,[u_2\,(u_1^2 - w_2^2 + u_3^2) + u_3\,(u_1^2 + u_2^2 -
 w_3^2)]\,\Gamma(0,0,-1;u_2+u_3,w_1,w_2+w_3)\,,
 \label{df_Pw1}
\end{eqnarray}
\end{widetext}
and where the two-electron integral $\Gamma$ is
\begin{eqnarray}
\Gamma(n_1,n_2,n_3,\alpha,\beta,\gamma) &\equiv& \int\frac{d^3r_1}{4\,\pi}\,\int\frac{d^3r_2}{4\,\pi}\,
\,r_1^{n_1-1}\,r_2^{n_2-1}\,r_{12}^{n_3-1} \nonumber \\ 
&& \times e^{-\alpha\,r_1-\beta\,r_2-\gamma\,r_{12}}\,,\label{Gamma}\\
\Gamma(0,0,-1,\alpha,\beta,\gamma) &=& \frac{1}{(\alpha-\beta)\,(\alpha+\beta)}
\ln\biggl(\frac{\gamma+\alpha}{\gamma+\beta}\biggr).
\end{eqnarray}
The recurrence relations for the integrals $g$
result from the differential equation (\ref{diff}) and can be written in a compact form as
\begin{eqnarray}
&&\sum^{n_1...n_6}_{i_1...i_6=0}\binom{n_1}{i_1}_{1/2}..\binom{n_6}{i_6}_{1/2}
\sigma(n_1-i_1,...,n_6-i_6)\, g(i_1,..,i_6) \nonumber \\
&& \hspace{3cm} = P(n_1-1,n_2,n_3,n_4,n_5,n_6), 
\label{df_rec}
\end{eqnarray}
where we use a Newton-like symbol notation
\begin{eqnarray} 
\binom{n}{0}_{1/2}=\frac{1}{2}, \quad \binom{n}{n}_{1/2}= 1,\nonumber \\
\quad \binom{n}{i}_{1/2} = \binom{n-1}{i}_{1/2} + \binom{n-1}{i-1}_{1/2},
\label{df_newt}
\end{eqnarray}
and
\begin{eqnarray}
P(n_1,n_2,n_3,n_4,n_5,n_6) &=& \nonumber \\ && \hspace{-2cm}
(-1)^{n_1 + \ldots + n_6} \frac{\partial^{n_1}}{\partial w_1^{n_1}} \ldots
\frac{\partial^{n_6}}{\partial u_3^{n_6}}\;P,  \\
\sigma(n_1,n_2,n_3,n_4,n_5,n_6) &=& \nonumber \\ && \hspace{-2cm} 
(-1)^{n_1 + \ldots + n_6} \frac{\partial^{n_1}}{\partial w_1^{n_1}} \ldots
\frac{\partial^{n_6}}{\partial u_3^{n_6}}\;\sigma.
\label{df_Pw1sig}
\end{eqnarray}
The relation (\ref{df_rec}) allows one to express the integral $g(n_1,..,n_6)$ 
with positive index $n_1$ through $g$-integrals with smaller nonnegative indices. 
Derivatives of polynomials in $P(n_1,n_2,n_3,n_4,n_5,n_6)$ and $\sigma(n_1,n_2,n_3,n_4,n_5,n_6)$ 
are calculated explicitly. The master integral $g(0,0,0,0,0,0)$ and the two-electron integrals 
$\Gamma$ are needed to start the evaluation of $g(n_1,n_2,n_3,n_4,n_5,n_6)$ from recursion relations. 
For $g(0,0,0,0,0,0)$ we implemented 
the formula of Fromm-Hill \cite{fromm} in the version improved by Harris \cite{harris_slater}. 
The calculation of the two-electron $\Gamma$ functions has been described in detail in Refs. 
\cite{sack_he,kor_he,harris_he}. In Eq. (\ref{df_rec}) the parameter
$w_1$ is distinguished on the right-hand side. We can define the same recurrence relations 
with other variables $w_a, u_a$ from expressions (\ref{df_Pw1}),
(\ref{df_rec}), and (\ref{df_Pw1sig}) by applying the tetrahedral symmetry.

The proposed recurrence scheme allows us to calculate integrals from higher
shells $\Omega$ very efficiently. 
In Table \ref{tb_ibasis} we present values in two reference points introduced by Fromm and Hill 
in Ref.~\cite{fromm}. These are the standard reference point (SRP) $w_a=u_a=1$ where $\sigma=-2$ and 
the auxiliary reference point (ARP) $w_a=1, u_a=0$ with $\sigma=1$.
Values for the last one can be compared to the known Hylleraas results. 
All presented digits are significant, which confirms the very good stability of the recursions 
at least at these reference points.
\begin{widetext}
\squeezetable
\begin{table*}[!thb]
\renewcommand{\arraystretch}{1.0}
\caption{Exponentially correlated integrals in SRP and ARP reference points}
\begin{ruledtabular}
\begin{tabular}{lw{1.40} w{1.40}}
                  &  \cent{{\rm ARP}:\; u_a=0,\;w_a=1} &  \cent{{\rm SRP}:\;u_a=w_a=1} \\ \hline 
$g(0,0,0,0,0,0)$    & 2.208\,310\,154\,388\,618\,874\,536\,424\,144\,0\;10^{-1} & 2.173\,757\,633\,275\,034\,284\,081\,325\,213\,9\;10^{-2}  \\
$g(1,0,0,0,0,0)$    & 2.208\,310\,154\,388\,618\,874\,536\,424\,144\,0\;10^{-1} & 1.086\,878\,816\,637\,517\,142\,040\,662\,607\,0\;10^{-2}  \\
$g(1,1,0,0,0,0)$    & 2.587\,328\,950\,655\,650\,145\,088\,210\,000\,9\;10^{-1} & 7.126\,296\,734\,072\,723\,436\,203\,022\,296\,4\,10^{-3}  \\
$g(2,0,0,0,0,0)$    & 3.658\,582\,716\,243\,175\,207\,969\,276\,574\,1\;10^{-1} & 8.571\,078\,153\,650\,184\,467\,645\,704\,204\,7\;10^{-3}  \\
$g(1,1,1,0,0,0)$    & 3.447\,454\,259\,102\,525\,282\,100\,115\,713\,9\;10^{-1} & 6.027\,278\,683\,541\,261\,617\,671\,441\,399\,9\;10^{-3}  \\
$g(3,0,0,0,0,0)$    & 8.803\,723\,087\,040\,150\,596\,505\,448\,579\,7\;10^{-1} & 9.316\,428\,522\,373\,859\,205\,286\,070\,304\,7\;10^{-3}  \\
$g(1,1,1,1,0,0)$    & 8.518\,518\,518\,518\,518\,518\,518\,518\,518\,5\;10^{0}  & 4.807\,529\,762\,877\,014\,687\,398\,150\,589\,3\;10^{-3}  \\
$g(4,0,0,0,0,0)$    & 2.849\,464\,173\,126\,685\,211\,199\,798\,289\,3\;10^{0}  & 1.296\,572\,573\,872\,689\,431\,790\,047\,714\,4\;10^{-2}  \\
$g(1,1,1,1,1,0)$    & 2.500\,000\,000\,000\,000\,000\,000\,000\,000\,0\;10^{0}  & 4.679\,985\,481\,131\,151\,729\,598\,237\,602\,7\;10^{-3}  \\
$g(5,0,0,0,0,0)$    & 1.176\,795\,411\,671\,425\,935\,279\,581\,659\,0\;10^{1}  & 2.205\,496\,365\,095\,795\,104\,480\,271\,164\,6\;10^{-2}  \\
$g(1,1,1,1,1,1)$    & 8.000\,000\,000\,000\,000\,000\,000\,000\,000\,0\;10^{0}  & 5.033\,124\,034\,041\,977\,640\,019\,585\,799\,2\;10^{-3}  \\
$g(6,0,0,0,0,0)$    & 5.962\,899\,567\,501\,152\,778\,486\,008\,087\,6\;10^{1}  & 4.441\,445\,631\,268\,318\,817\,661\,363\,268\,6\;10^{-2}  \\
$g(2,2,2,2,2,2)$    & 4.761\,568\,658\,224\,658\,953\,775\,935\,886\,3\;10^{4}  & 1.467\,676\,925\,382\,090\,748\,693\,886\,720\,7\;10^{-1}  \\
$g(12,0,0,0,0,0)$   & 1.998\,667\,989\,835\,473\,493\,169\,581\,298\,3\;10^{7}  & 3.793\,674\,032\,518\,970\,385\,528\,210\,586\,4\;10^{1}   \\
$g(3,3,3,3,3,3)$    & 4.595\,097\,600\,000\,000\,000\,000\,000\,000\,0\;10^{9}  & 5.979\,770\,415\,024\,714\,287\,528\,834\,619\,9\;10^{1}   \\
$g(18,0,0,0,0,0)$   & 1.779\,533\,879\,212\,729\,823\,179\,024\,201\,1\;10^{14} & 6.240\,107\,562\,655\,131\,418\,896\,693\,018\,6\;10^{5}   \\
\end{tabular}
\end{ruledtabular}
\label{tb_ibasis}
\end{table*}
\end{widetext}

\begin{widetext}
\squeezetable
\begin{table*}[bth]
\renewcommand{\arraystretch}{1.0}
\caption{Numerical values of  $g(1,1,1,1,1,1)$ around the singular point $\sigma=0$: $w_{1,2,3}=1$, $u_{1,2,3}=\alpha$.}
\begin{ruledtabular}
\begin{tabular}{ w{1.30}l w{4.35}}
\cent{\alpha}              & \cent{\sigma} & \cent{g(1,1,1,1,1,1)} \\ \hline 
0.577\,3                 & $ 1.741\,3\;10^{-4}$                     & 4.715\,168\,989\,550\,199\,879\,343\,821\,168\;10^{-2}  \\
0.577\,350\,2            & $ 2.396\,798\,8\;10^{-7}$                & 4.713\,684\,016\,763\,871\,477\,743\,039\,334\;10^{-2}  \\
0.577\,350\,269          & $ 6.568\,829\,17\;10^{-10}$              & 4.713\,681\,976\,034\,959\,469\,113\,297\,912\;10^{-2}  \\
0.577\,350\,269\,189\,6  & $ 8.925\,087\,775\,552\;10^{-14}$        & 4.713\,681\,970\,427\,392\,724\,217\,475\,890\,10^{-2}  \\
0.577\,350\,269\,189\,625& $ 2.648\,337\,377\,078\,125\;10^{-15}$   & 4.713\,681\,970\,426\,653\,329\,868\,706\,104\;10^{-2}  \\
0.577\,350\,269\,189\,625\,764\,5& $ 3.169\,230\,531\,337\,161\,925\;10^{-20}$     & 4.713\,681\,970\,426\,630\,719\,189\,520\,726\;10^{-2}  \\
0.577\,350\,269\,189\,625\,764\,509\,148\,7& $ 2.788\,669\,608\,438\,596\,620\,038\,649\,3\;10^{-25}$ & 4.713\,681\,970\,426\,630\,718\,918\,940\,842\;10^{-2}  \\
3.^{-1/2}                & $ 0 $                                                   & 4.713\,681\,970\,426\,630\,718\,918\,938\,462\;10^{-2}  \\
0.577\,350\,269\,189\,625\,764\,509\,148\,8& $ -6.754\,320\,066\,991\,579\,670\,162\,432\;10^{-26}$   & 4.713\,681\,970\,426\,630\,718\,918\,937\,885\;10^{-2}  \\
0.577\,350\,269\,189\,625\,764\,6& $-3.147\,178\,562\,004\,038\,394\,8 \;10^{-19}$ & 4.713\,681\,970\,426\,630\,716\,231\,943\,331\;10^{-2}  \\
0.577\,350\,269\,189\,626& $-8.157\,642\,380\,596\,28\;10^{-16}$    & 4.713\,681\,970\,426\,623\,754\,094\,755\,315\;10^{-2}  \\
0.577\,350\,269\,189\,7  & $-2.571\,592\,837\,582\,7\;10^{-13}$     & 4.713\,681\,970\,424\,435\,146\,822\,397\,542\;10^{-2}  \\
0.577\,350\,270          & $-2.807\,218\,7\;10^{-9}$                & 4.713\,681\,946\,459\,185\,584\,378\,695\,888;10^{-2}  \\
0.577\,350\,3            & $-1.067\,302\,7\;10^{-7}$                & 4.713\,681\,059\,186\,068\,003\,930\,086\,238\;10^{-2}  \\
0.577\,4                 & $-1.722\,8\;10^{-4}$                     & 4.712\,211\,406\,354\,297\,107\,821\,210\,234\;10^{-2}  \\
\end{tabular}
\end{ruledtabular}
\label{tb_ibsing}
\end{table*}
\end{widetext}
The integrals $g(n_1,n_2,n_3,n_4,n_5,n_6)$ with $\sigma$ is close to zero are difficult 
to evaluate because the recurrence relations are numerically unstable. 
Following Harris' studies on the master integral $g(0,0,0,0,0,0)$ in
Ref.~\cite{harris_slater}, we considered $w_a=1, u_a=\alpha$ close 
to $\alpha=3^{-1/2}$ where $\sigma$ is exactly equal to zero. 
As an example we present in Table~\ref{tb_ibsing} values for $g(1,1,1,1,1,1)$ 
which needs six evaluations of  recursions (\ref{df_rec}).
Close to the critical point $\sigma=0$, we used Bailey's multiprecision  library
\cite{bailey_lib}. We control the number of significant digits 
by dynamical estimating and adjusting the appropriate precision of the arithmetic. 
As presented in Table \ref{tb_ibsing}, we can approach the critical point $\sigma=0$
as close as we need for practical purposes. This strategy of course slows down 
the algorithm significantly, but in practical applications the parameters close to
the critical point $\sigma=0$ are very rare. This strategy of controlling
precision in the region of instabilities allows one to cross $\sigma=0$ points 
in the minimization of the nonrelativistic energy. In Table~\ref{tb_nenergy} 
we present results for nonrelativistic energies for the ground state of Li 
obtained with global minimization of all nonlinear parameters in the basis 
length of $N=$ 128, 256, 512 respectively.
\begin{table}[!thb]
\renewcommand{\arraystretch}{1.0}
\caption{Nonrelativistic energy of the Li ground state using exponentially correlated basis set
with the number of functions $N$, $\infty$ is the extrapolated value from Ref.~\cite{lit_rel}
using Hylleraas basis functions. }
\begin{tabular}{lw{6.19}}
$N $               &  \cent{{\cal E}^{(2)}} \\ \hline 
32   & -7.478\,059\,40  \\
64   & -7.478\,060\,050  \\
128  & -7.478\,060\,304\,6  \\
256  & -7.478\,060\,321\,31  \\
512  & -7.478\,060\,323\,427  \\
$\infty$& -7.478\,060\,323\,910\,1(3)\\ \hline
\end{tabular}
\label{tb_nenergy}
\end{table}
The achieved precision is much higher than that from similar number of Hylleraas functions.

\subsection{Extended integrals}
In this section we present an algorithm for calculations of extended integrals with 
$1/r_a^2$ or $1/r_{ab}^2$ factors in Eq~(\ref{df_g}). This means that some of
indices in $g(n_1,n_2,n_3,n_4,n_5,n_6)$ are equal to $-1$. 
Fully correlated exponent in Eq.~(\ref{df_g}) gives the opportunity to 
obtain extended integrals by using either a single integration over $w_a$ or $u_a$ i.e.
\begin{eqnarray}
g(n_1,n_2,n_3,-1,n_5,n_6) &=& \nonumber \\ && \hspace{-1.5cm}
\int_{u_1}^\infty {\rm d} u_1\,g(n_1,n_2,n_3,0,n_5,n_6)\,,\nonumber \\
\label{df_gext1}
\end{eqnarray}
or a double integration i.e. 
\begin{eqnarray}
g(n_1,-1,n_3,-1,n_5,n_6) &=& \nonumber \\ && \hspace{-1.5cm}
\int_{u_1}^\infty {\rm d} \, u_1 \int_{w_2}^\infty {\rm d} w_2 \, g(n_1,0,n_3,0,n_5,n_6). \nonumber \\
\label{df_gext2}
\end{eqnarray}
The adaptive increase of the arithmetic precision close to critical points
$\sigma=0$ is necessary here for the precise evaluation of
$g(n_1,n_2,n_3,n_4,n_5,n_6)$. 
Moreover, 
one is able to perform accurately this integration by using $N$-point generalized Gaussian 
quadrature with logarithmic end-point singularity \cite{rokhlin}
\begin{eqnarray} 
I &=& \int_0^1 dx\,\bigl[W_1(x) + \ln(x)\,W_2(x)\bigr] \nonumber \\
  &\approx& \sum_{i=1}^N\,w_i\,\bigl[W_1(x_i) + \ln(x_i)\,W_2(x_i)\bigr]\,,
\label{df_rokhlin}
\end{eqnarray}
where $W_i$ are regular functions on the interval $(0,1)$.
This quadrature becomes exact for $W_i$ being polynomials of maximal degree
$N-1$, and the example of 30 point quadrature, which is used
through out this paper for calculation of mean values, is presented in
Appendix A.

If integration variable $u$ in Eq. (\ref{df_gext1}) is mapped into the interval
$(0,1)$ by the following change of variable
\begin{equation}
\int_0^\infty du\,f(u) = \int_0^1 dx\,\frac{1}{x^2}\,f\biggl(\frac{1}{x}-1\biggr) \label{1fold}
\end{equation}
then the quadrature with logarithmic end-point
singularity is sufficient for one-dimensional integrals in Eq. (\ref{df_gext1}), 
where $N=30$ quadrature allows  one to obtain about 30 significant digits 
as shown in Table~\ref{tb_gm}. 
\begin{widetext}
\squeezetable
\begin{table*}[!thb]
\renewcommand{\arraystretch}{1.0}
\caption{Examples of extended integrals with $1/r_{23}^2$ in SRP and ARP
  reference points  calculated using numerical integration with $N=30$.}
\begin{ruledtabular}
\begin{tabular}{lw{1.38} w{1.38}}
&  \cent{{\rm ARP}:\; u_a=0,\; w_a=1} &  \cent{{\rm SRP}:\; u_a=w_a=1} \\ \hline 
$g(0,0,0,-1,0,0$)   & 3.852\,610\,933\,969\,379\,240\,110\,048\,369\,9 \;10^{-1} & 9.496\,501\,144\,947\,432\,180\,784\,237\,909\,6 \;10^{-2}  \\
$g(1,0,0,-1,0,0)$   & 3.027\,449\,106\,575\,050\,367\,308\,702\,234\,4 \;10^{-1} & 3.729\,972\,160\,750\,497\,052\,308\,263\,021\,7 \;10^{-2}  \\
$g(1,1,0,-1,0,0)$   & 2.360\,700\,062\,552\,231\,395\,696\,277\,796\,4 \;10^{-1} & 1.895\,953\,170\,412\,559\,215\,217\,374\,496\,4 \;10^{-2}  \\
$g(2,0,0,-1,0,0$)   & 4.360\,947\,194\,620\,688\,310\,533\,551\,110\,5 \;10^{-1} & 2.519\,224\,983\,963\,737\,154\,014\,628\,472\,2 \;10^{-2}  \\
$g(1,1,1,-1,0,0)$   & 2.503\,315\,976\,630\,860\,793\,316\,473\,181\,3 \;10^{-1} & 1.453\,670\,463\,091\,034\,929\,350\,269\,890\,4 \;10^{-2}  \\
$g(3,0,0,-1,0,0$)   & 9.678\,534\,014\,755\,149\,691\,745\,602\,065\,7 \;10^{-1} & 2.452\,489\,069\,429\,084\,805\,380\,219\,451\,1 \;10^{-2}  \\
$g(1,1,1,-1,1,0)$   & 4.965\,970\,761\,362\,395\,801\,363\,888\,667\,7 \;10^{-1} & 8.929\,779\,730\,282\,261\,688\,941\,298\,690\,9 \;10^{-3}  \\
$g(4,0,0,-1,0,0)$   & 2.994\,734\,857\,888\,708\,511\,724\,418\,787\,7 \;10^{0}  & 3.140\,438\,234\,282\,031\,347\,870\,179\,278\,8 \;10^{-2}  \\
$g(1,1,1,-1,1,1$)   & 1.333\,333\,333\,333\,333\,333\,333\,333\,333\,3 \;10^{0}  & 7.833\,365\,875\,543\,837\,545\,344\,094\,254\,1 \;10^{-3}  \\
$g(5,0,0,-1,0,0$)   & 1.207\,413\,986\,734\,661\,793\,622\,502\,357\,9 \;10^{1}  & 5.003\,447\,967\,255\,362\,321\,317\,326\,500\,6 \;10^{-2}  \\
$g(2,2,2,-1,2,2)$   & 5.436\,536\,048\,634\,697\,021\,325\,813\,246\,7 \;10^{2}  & 6.776\,488\,986\,538\,662\,624\,450\,203\,053\,8\;10^{-2}  \\
$g(10,0,0,-1,0,0)$  & 1.820\,037\,338\,296\,110\,774\,925\,172\,748\,9 \;10^{5}  & 4.998\,826\,524\,680\,257\,565\,021\,055\,520\,2 \;10^{0}  \\
$g(3,3,3,-1,3,3)$   & 2.636\,015\,376\,623\,376\,623\,376\,623\,376\,6 \;10^{6}  & 6.615\,220\,449\,456\,842\,761\,016\,682\,847\,5 \;10^{0}  \\
$g(15,0,0,-1,0,0)$  & 4.364\,311\,343\,749\,328\,364\,240\,301\,511\,7 \;10^{10} & 6.183\,347\,201\,330\,971\,024\,067\,326\,468\,8 \;10^{3} 
\end{tabular}
\end{ruledtabular}
\label{tb_gm}
\end{table*}
\end{widetext}
The presented values for integrals are obtained 
at ARP and SRP points. Some of them can be found in the literature, for example 
$g(2,2,2,-1,2,2)$ at ARP point \cite{pelzl_ext,rec_ext}, which corresponds to 
a Hylleraas type of integral. Perfect agreement with those results,
demonstrates  high accuracy is achieved for the extended integrals.
The proposed evaluation method fully relies on properties of the 
recurrence algorithm for basis integrals, which must be very stable 
on the integration path over the corresponding parameter.

In comparison to the one-dimensional integral in Eq. (\ref{df_gext1}), 
the convergence of two-dimensional integral Eq.~(\ref{df_gext2}) with respect
to the number of integration points is much worse. For this reason
we use a slightly different mapping into $(0,1)$ intervals, which is
\begin{eqnarray}
\int_0^\infty du\,\int_0^\infty dw\,f(u,w) &=& \nonumber \\ &&\hspace{-20ex} 
\int_0^1 dx\,\int_0^1 dy\,\frac{4}{x^3\,y^3}\,f\biggl(\frac{1}{x^2}-1,\frac{1}{y^2}-1\biggr). \label{2fold}
\end{eqnarray}
The numerical convergence of the integral in Eq. (\ref{df_gext2}) is the worst 
for the case $n_3=0$, where the leading asymptotics includes a square of the logarithm.  
For $n_3>0$ convergence improves significantly. 
The use of Gaussian quadrature adapted to
logarithmic end-point singularity  with 30 points is enough for practical applications. 
In Table~\ref{tb_gmm} we
presented numerical values for $g(n_1,-1,n_3,-1,n_5,n_6)$ in ARP and SRP reference points 
with the accuracy of $10^{-16}$. In the case of $n_3=0$ they have been
obtained with 60 point quadrature. It is possible to obtain even  higher accuracy 
for $n_3>0$,  but further improvement for $n_3=0$ 
requires a more sophisticated integration strategy. There are no such
problems with integration involving parameters which are attached
to opposite edges of the tetrahedron, i.e. $g(-1,n_2,n_3,-1,n_5,n_6)$, 
see Fig.~\ref{fig_tetrahedron}, so this case of integral in Eq. (\ref{df_gext1})
with $n_3=0$ is the one which limits accuracy of mean values.
\begin{widetext}
\begin{table*}[!thb]
\renewcommand{\arraystretch}{1.0}
\caption{ Examples of extended integrals $g(n_1,-1,n_3,-1,n_5,n_6)$ in SRP and
  ARP reference points, numerical quadrature with $N=60$ points, all digits
  are significant}
\begin{ruledtabular}
\begin{tabular}{lw{1.23} w{1.23}}  &  \cent{{\rm ARP}:\; u_a=0,\; w_a=1} & \cent{{\rm SRP}:\; u_a=w_a=1} \\ \hline
$g(0,-1,0,-1,0,0)$   &  1.884\,392\,088\,158\,216\;10^{0} & 8.424\,892\,130\,134\,382\;10^{-1}  \\
$g(1,-1,0,-1,0,0)$   &  9.485\,660\,506\,739\,961\;10^{-1}& 1.797\,709\,314\,546\,008\;10^{-1}  \\
$g(1,-1,1,-1,0,0)$   &  4.284\,596\,512\,743\,028\;10^{-1}& 4.614\,367\,395\,977\,269\;10^{-2}  \\
$g(2,-1,0,-1,0,0)$   &  1.165\,927\,539\,416\,184\;10^{0} & 8.712\,176\,344\,048\,701\;10^{-2}  \\
$g(1,-1,1,-1,1,0)$   &  6.829\,291\,358\,121\,588\;10^{-1}& 2.470\,684\,625\,918\,397\;10^{-2}  \\
$g(3,-1,0,-1,0,0)$   &  2.440\,178\,087\,422\,623\;10^{0} & 6.789\,663\,816\,285\,613\;10^{-2}  \\
$g(1,-1,1,-1,1,1)$   &  1.467\,401\,100\,272\,340\;10^{0} & 1.868\,910\,909\,646\,633\;10^{-2}  \\
$g(4,-1,0,-1,0,0)$   &  7.400\,703\,871\,489\,098\;10^{0} & 7.375\,782\,376\,091\,605\;10^{-2}  \\
$g(2,-1,2,-1,2,2)$   &  1.325\,932\,535\,285\,045\;10^{2} & 3.788\,534\,374\,425\,215\;10^{-2}  \\
$g(7,-1,0,-1,0,0)$   &  8.898\,402\,193\,692\,669\;10^{2} & 3.628\,968\,910\,010\,360\;10^{-1}  \\
\end{tabular}
\end{ruledtabular}
\label{tb_gmm}
\end{table*}
\end{widetext}

\subsection{Expectation values}
The basis class of integrals ($n_i\geq0$) and the class with the one index equal to $-1$ is sufficient for 
all mean values of operators like those in the Breit-Pauli Hamiltonian, 
Eq.~(\ref{df_H4}). As an example we demonstrate their evaluation for the lithium ground state. 
In Table~\ref{tb_rel} we present results for 
the Dirac $\delta$ functions using the Drachman formulae \cite{Dra81}
\begin{eqnarray}
\langle 4 \pi \delta^3(r_{a}) \rangle &=&
  \left\langle \frac{4}{r_{a}}(E_0-V)\right\rangle - \sum_c\left\langle{
\vec \nabla}_c\phi\left|\frac{2}{r_{a}}\right|{\vec \nabla}_c\phi\right\rangle\,,
\nonumber \\  \\
 \langle 4\pi \delta^3(r_{ab}) \rangle &=&
  \left\langle \frac{2}{r_{ab}}(E_0-V)\right\rangle -\sum_c\left\langle{
\vec \nabla}_c\phi\left|\frac{1}{r_{ab}}\right|{\vec \nabla}_c\phi\right\rangle,
\nonumber \\
\end{eqnarray}
where $V$ is a total interaction potential. The similar prescriptions can be used for $p_a^4$ operator 
\begin{equation}
\Bigl\langle \sum_a p_a^4 \Bigr\rangle =  4 \langle (E_0-V)^2 \rangle -\sum_{b>c} \langle
\nabla_b^2 \phi | \nabla_c^2 \phi \rangle.
\end{equation}
These forms significantly improve accuracy of numerical results 
in comparison to direct calculations, like those presented in our previous paper \cite{slater1},
see Table ~\ref{tb_rel}.
Nevertheless, with 256 functions they are about two digits less accurate
than the most precise results obtained from 9564 Hylleraas basis functions.
These Hyleraas results are slightly more accurate than those in
\cite{lit_wave} due to better optimization of the nonrelativistic wave function.
 
As we have noticed, for the nonrelativistic 
energy one needs approximately six times smaller basis set of exponentially 
correlated functions as compared to Hylleraas functions to obtain similar accuracy,
and the same is confirmed for the mean values of operators. 
The achieved accuracy is limited 
only by the number of basis functions, which nevertheless should be well optimized. 
In practice it demands more computing power than 
we used up to now and a parallel version of the algorithm would be necessary
for optimization of a large number of Slater functions. 
\begin{widetext}
\squeezetable
\begin{table*}[!hbt]
\renewcommand{\arraystretch}{1.0}
\caption{ Expectation value for Breit-Pauli operators for the ground state
  involving integrals with one $n_i=-1$. Implicit summation over $a$, or over
  pairs $a>b$ is assumed. Last entries are extrapolated results obtained in Hylleraas
  basis set}
\begin{ruledtabular}
\begin{tabular}{lw{1.11}w{1.12}w{2.11}w{1.12}w{3.8}w{1.12}}
\cent{N}  & \cent{r_a^{-2}} &  \cent{r_{ab}^{-2}} & \cent{\delta^3(r_a)} & \cent{\delta^3(r_{ab})} & \cent{p_a^4} & 
\cent{p_a^i\,r_{ab}^{-3}(\delta^{ij} r_{ab}^2 + r_{ab}^i r_{ab}^j) p_b^j}
\\ \hline \hline
1  & 30.082\,797\,986\,7 & 4.506\,456\,504\,1 & 13.764\,569\,952\,4  & 0.536\,449\,208\,3 & 625.582\,840\,9 & 0.936\,654\,826 \\
2  & 29.747\,608\,655\,0 & 4.576\,769\,565\,1 & 13.585\,859\,628\,1  & 0.561\,647\,862\,3 & 611.553\,996\,7 & 1.138\,521\,450 \\
4  & 30.130\,068\,846\,4 & 4.443\,919\,421\,0 & 13.787\,803\,129\,0  & 0.543\,639\,175\,1 & 625.468\,724\,2 & 0.914\,892\,317 \\
8  & 30.187\,481\,732\,0 & 4.421\,446\,718\,9 & 13.815\,726\,640\,6  & 0.543\,582\,099\,9 & 627.407\,849\,4 & 0.903\,179\,423 \\
16 & 30.241\,078\,773\,0 & 4.381\,681\,063\,4 & 13.842\,637\,278\,6  & 0.544\,318\,958\,8 & 628.479\,718\,2 & 0.871\,711\,023 \\
32 & 30.240\,966\,286\,8 & 4.381\,283\,593\,8 & 13.842\,598\,063\,3  & 0.544\,325\,804\,3 & 628.457\,736\,5 & 0.871\,331\,224 \\
64 & 30.240\,892\,554\,9 & 4.381\,232\,031\,6 & 13.842\,567\,782\,2  & 0.544\,325\,359\,5 & 628.451\,398\,8 & 0.871\,268\,418 \\
128& 30.240\,987\,196\,9 & 4.381\,186\,567\,3 & 13.842\,617\,080\,2  & 0.544\,324\,836\,8 & 628.450\,904\,8 & 0.871\,208\,043 \\
256& 30.240\,973\,605\,8 & 4.381\,176\,947\,3 & 13.842\,611\,088\,3  & 0.544\,324\,684\,9 & 628.449\,069\,5 & 0.871\,196\,220 \\ \\
Hyll. & 30.240\,972\,72(3) & 4.381\,176\,64(4) & 13.842\,610\,86(3) & 0.544\,324\,632\,5(7)   & 628.448\,985(12) & 0.871\,195\,62 (14)
\end{tabular}
\end{ruledtabular}
\label{tb_rel}
\end{table*}
\end{widetext}
In Table \ref{tb_opgmm1} we present numerical 
values for typical operators in higher order perturbation theory i.e. 
$m \alpha^6$ correction to the energy. For these operators we need to use all 
the discussed classes of integrals. 
$g(n_1,n_2,n_3,n_4,n_5,n_6)$ integrals with two parameters
equal to $-1$ are obtained with double integration with $30 \times 30$ points
and all the presented digits in Table \ref{tb_opgmm1} are accurate
for the corresponding approximate wave function. 
\begin{widetext}
\begin{table*}[!thb]
\renewcommand{\arraystretch}{1.0}
\caption{ Expectation value for operators for operators involving extended class of integrals.}
\begin{ruledtabular}
\begin{tabular}{lw{4.7}w{4.7}w{2.9}w{3.8}w{4.7}}
\cent{N} & \cent{\sum_{a>b}r_a^{-2} r_b^{-2}} 
         & \cent{\sum_{a\neq b} r_{a}^{-2} r_{ab}^{-2}} 
         & \cent{\sum_{a>b>c}r_{ab}^{-2} r_{bc}^{-2}} 
         & \cent{\sum_{a\neq b}\vec r_{a}\cdot\vec r_{ab} r_{a}^{-3} r_{ab}^{-3}} 
         & \cent{\sum_a\,\sum_{b>c}p_b^2 r_a^{-1} p_c^2}
\\ \hline 
1  & 204.916\,879 & 289.283\,997 & 1.177\,470\,80 & 45.945\,338\,1 & 842.720\,739 \\
2  & 211.038\,011 & 297.213\,636 & 1.112\,887\,76 & 48.080\,149\,7 & 891.514\,656 \\
4  & 202.558\,151 & 284.360\,921 & 1.071\,904\,15 & 45.386\,279\,3 & 840.040\,129 \\
8  & 202.665\,783 & 283.643\,499 & 1.066\,132\,65 & 45.443\,462\,5 & 844.843\,155\\
16 & 202.840\,737 & 282.053\,625 & 1.064\,131\,54 & 45.240\,730\,3 & 853.983\,531\\
32 & 202.852\,464 & 282.008\,017 & 1.064\,188\,66 & 45.236\,709\,9 & 854.271\,406\\
64 & 202.862\,382 & 282.005\,597 & 1.064\,179\,55 & 45.236\,342\,7 & 854.374\,209\\
128& 202.883\,252 & 282.002\,183 & 1.064\,178\,17 & 45.236\,047\,4 & 854.554\,104\\
\end{tabular}
\end{ruledtabular}
\label{tb_opgmm1}
\end{table*}
\end{widetext}

\section{Summary}

Our primary motivation for developing explicitly correlated exponential 
basis set is the efficient representation of the wave function in a
small number of basis functions. We applied it for the accurate numerical
calculation of expectation values of some operators corresponding to higher order
relativistic and QED effects. They involve integrals with quadratic inverse
powers of at least two interparticle distances, which are the most difficult
in the evaluation. Using this compact and very flexible 
correlated exponential basis set, we are aiming to determine $m\,\alpha^6$ 
and $m\,\alpha^7$ effects in the hyperfine and fine structure of lithium-like
systems, which have not been investigated so far.

\section*{Acknowledgments}
This work was supported by NIST through Precision Measurement Grant
PMG 60NANB7D6153.

\appendix
\renewcommand{\theequation}{\Alph{section}\arabic{equation}}
\renewcommand{\thesection}{\Alph{section}}

\section{Weights and nodes of generalized Gaussian quadrature with logarithmic end-point singularity}

We present here a set of 30 nodes and corresponding weights 
for generalized Gaussian quadrature 
with logarithmic end-point singularity \cite{rokhlin},
which was obtained using the algorithm described in Appendix A of Ref. \cite{lit_wave}.
We used these quadrature points for numerical integration of extended integrals in 
Tables \ref{tb_gm}, \ref{tb_rel}, and \ref{tb_opgmm1}.
\begin{widetext}
\squeezetable
\begin{table*}[!hbt]
\renewcommand{\arraystretch}{1.0}
\caption{ Weights and nodes of generalized Gaussian quadrature with logarithmic end-point singularity}
\begin{ruledtabular}
\begin{tabular}{lw{2.38}w{1.38}}
\cent{N}  & \cent{\rm nodes} & \cent{\rm weights}
\\ \hline 
1    &7.323\,797\,44272605707927651215334608\, 10^{-6}  &2.79892154309547416710987828334736\, 10^{-5}    \\
2    &1.100\,447\,00457774879623368108247943\, 10^{-4}  &2.17365526502541548589108155033626\, 10^{-4}    \\
3    &5.469\,183\,26183967432457719790539051\, 10^{-4}  &7.20703586534389568501488362593518\, 10^{-4}    \\
4    &1.701\,857\,51910164118701225273639596\, 10^{-3}  &1.67446096505498972978224436308689\, 10^{-3}    \\
5    &4.083\,863\,60971437462932298732108869\, 10^{-3}  &3.19128240641146524350794278296664\, 10^{-3}    \\
6    &8.300\,041\,17688233905931162736411746\, 10^{-3}  &5.35378831352933564469099803155786\, 10^{-3}    \\
7    &1.502\,297\,81560799959114892646895942\, 10^{-2}  &8.20962136808195567645775658689256\, 10^{-3}    \\
8    &2.495\,392\,36157545503008513656313159\, 10^{-2}  &1.17680292130848124961801044296255\, 10^{-2}    \\
9    &3.878\,338\,61710629356474590553368630\, 10^{-2}  &1.59981435048914024332333704241883\, 10^{-2}    \\
10   &5.715\,089\,84811763898261154101387036\, 10^{-2}  &2.08290410936242947283573768810863\, 10^{-2}    \\
11   &8.060\,574\,14726551690255082886279043\, 10^{-2}  &2.61515976613093133103470957985343\, 10^{-2}    \\
12   &1.095\,703\,94234517942760556318344690\, 10^{-1}  &3.18220682694563815827694187138199\, 10^{-2}    \\
13   &1.443\,083\,73001500812672361144038109\, 10^{-1}  &3.76672559998067174250314484868819\, 10^{-2}    \\
14   &1.848\,979\,49427531958876501649047717\, 10^{-1}  &4.34910622632233969988331688489985\, 10^{-2}    \\
15   &2.312\,130\,01548255181588787651651928\, 10^{-1}  &4.90821532284030403760822595772833\, 10^{-2}    \\
16   &2.829\,119\,60197207748156616199457019\, 10^{-1}  &5.42224286612626157484616606008227\, 10^{-2}    \\
17   &3.394\,354\,81190852660583946762696458\, 10^{-1}  &5.86959442909769135605447275253662\, 10^{-2}    \\
18   &4.000\,131\,13109450190847537317246366\, 10^{-1}  &6.22979181149364796169604799658096\, 10^{-2}    \\
19   &4.636\,788\,56939306293165441422602120\, 10^{-1}  &6.48434457072330546001338183346875\, 10^{-2}    \\
20   &5.292\,951\,42741036253816689241764505\, 10^{-1}  &6.61755598512543778100952715354295\, 10^{-2}    \\
21   &5.955\,843\,95325645009469243558808251\, 10^{-1}  &6.61722952772004782470943263263082\, 10^{-2}    \\
22   &6.611\,670\,40396915198682135573185723\, 10^{-1}  &6.47524589229209230876914855177978\, 10^{-2}    \\
23   &7.246\,045\,28176032770586180257340226\, 10^{-1}  &6.18798583499546375301194127428597\, 10^{-2}    \\
24   &7.844\,457\,34734941543894377419518002\, 10^{-1}  &5.75658036634420432925717153926242\, 10^{-2}    \\
25   &8.392\,749\,51478663018317278487334798\, 10^{-1}  &5.18697691924656775668123741331021\, 10^{-2}    \\
26   &8.877\,595\,97617343115537224494301863\, 10^{-1}  &4.48981784955672218807632566389701\, 10^{-2}    \\
27   &9.286\,957\,95963227580314671288014994\, 10^{-1}  &3.68013625003694933496657925550436\, 10^{-2}    \\
28   &9.610\,500\,59187409711054297148665964\, 10^{-1}  &2.77688703774135920873618253138605\, 10^{-2}    \\
29   &9.839\,957\,03521288981120560285698734\, 10^{-1}  &1.80238737841607431150476240903211\, 10^{-2}    \\
30   &9.969\,459\,58679763051044061968242492\, 10^{-1}  &7.82767019549675700264134910161448\, 10^{-3}    \\
\end{tabular}
\end{ruledtabular}
\label{tb_quadrature}
\end{table*}
\end{widetext}
\end{document}